# Discussion on phase ambiguity and multiple beam generation in coherent beam combining system

Haolong Jia, [1,3,4,†] Jing Zuo, [2,3,4,†] Qiliang Bao, [1,3,4,*] Chao Geng, [2,3,4,*,*] Ao Tang, [2,3,4] Yihan Luo, [1,3,4] Ziqiang Li, [2,3,4] Jing Jiang, [1,3,4] Feng Li, [2,3] Fan Zou, [2,3,4] Xu Yang, [2,3,4,5] Ziting Pan, [2,3,4] Jiali Jiang, [2,3] Jianpeng Ren, [6] and Xinyang Li, [2,3,4]

[1] *Key Laboratory of Optical Engineering, Chinese Academy of Sciences, Chengdu 610209, China*

[2] *Key Laboratory on Adaptive Optics, Chinese Academy of Sciences, Chengdu 610209, China*

[3] *Institute of Optics and Electronics, Chinese Academy of Sciences, Chengdu 610209, China*

[4] *University of Chinese Academy of Sciences, Beijing 100049, China*

[5] *University of Electronic Science and Technology of China, Chengdu 611731, China*

[6] *Chengdu Boltzmann Intelligence Technology Co., Ltd, Chengdu 610094, China*

*Corresponding author: 13880972802@163.com; ** corresponding author: blast_4006@126.com*

**Abstract:** There exists the phase ambiguity problem in the coherent beam combining (CBC) system with centrosymmetric arrays, which means that multiple different piston aberrations may generate the same far-field image. This will cause that the far-field image can not correctly reflect the phase information, resulting in the performance degradation of image-based intelligent algorithms. In this paper, we make a theoretical analysis on phase ambiguity. To the best of our knowledge, we give the number and descriptions of all solutions of the phase ambiguity problem in above system for the first time. A method to solve phase ambiguity is proposed, which requires no additional optical devices. We designed simulations to verify our conclusions and methods. We believe that our work solves the phase ambiguity problem in theory and is conducive to improving the performance of image-based algorithms. In addition, we designed a two-stage algorithm to generate Bi-beam, which have valuables application in laser propagation.

## 1. Introduction

High-power lasers are widely applied in many fields, including Lidar system, Space Communication, Laser Medicine, Material Processing and so on [1-6]. Therefore, obtaining high-power lasers has important practical value. According to the theory of wave optics, when the wavelength of incident light is fixed, we should increase the apertures in optical systems aiming at improving the laser power. However, the high processing cost and manufacturing difficulty of large aperture optical system limit its development and application. Fortunately, coherent beam combining (CBC) provides a way to obtain high-power laser through multiple small apertures [1, 2, 7-9]. In 2021, Civan Laser achieved high-power and shape controllable beam based on CBC and optical phased array (OPA). The results show that the co-phase output in CBC system can achieve better power improvement than the series amplifiers. The core challenge of achieving co-phase output is to correct and lock the piston aberration quickly between each subbeam [10-17]. In the past, researchers usually used stochastic parallel gradient descent (SPGD) to compensate the piston aberration [16, 17]. This blind optimization algorithm costs a number of iterative steps before reaching convergence, which limits its application in practice. The main reason for the inefficient iteration is the insufficient utilization of system-state information. In SPGD, researchers only take the power in the bucket (PIB) as the objective function for optimization, ignoring other valuable information that can reflect the piston aberration, such as light intensity distribution. At the same time, the iterative process is nonheuristic, which means it is difficult for the algorithm to select an appropriate optimization strategy according to the current state.

With the development of machine learning, various intelligent algorithms are introduced into CBC [18-21]. In 2019, Hou *et al.* predicted the rough value of piston aberration according to the far-field image at the defocus plane with convolutional neural network (CNN), and then applied SPGD based on the prediction, so as to increase the convergence speed [18]. In 2020, Liu *et al.* employ CNN to measure the beam-pointing and piston aberration of subbeams from the far-field image in a two-beam coherent beam combining system [19]. In 2021, Henrik *et al.* applied deep reinforcement learning to CBC [20]. In this method, the neural network receives the near-field image as state information, and obtains the optimization strategy through reinforcement learning. In this method, the trained agent takes the far-field image as input, and then controls the input piston aberration through reinforcement learning. These studies indicate that the introduction of image information can effectively improve the convergence-performance of the algorithm, which would better meet the needs of practical applications.

However, researchers find that phase ambiguity exists in centrosymmetric arrays [22, 23], which means different piston aberration may generate the same far-field image. In this case, the far-field image can not provide sufficient information of the current piston aberration, leading to the gain of introducing image information reducing. If we can solve the phase ambiguity problem, the introduced image information will accurately reflect the piston aberration, which helps for high-speed and accurate prediction.

CBC can be extended to many applications. One of the valuable applications is that we could control the phase to generate multiple beams after realizing co-phase, which can be applied in Laser Transmission, Material Processing and other fields [24-27]. The challenge is to find the specific phase of generating multiple beams in unknown arrays. If we can design an appropriate algorithm to solve this problem, we can replace the existing mechanical method with phase-control method, which is faster, more accurate and more flexible.

The main contributions of this paper are as follows:
(1) We prove that the phase ambiguity may occur in centrosymmetric-array CBC system. Through mathematical derivation, we determine all the solutions, and give their mathematical description.
(2) We provide a method to solve the phase ambiguity problem without additional optical devices.
(3) In addition to theoretical proof, we designed simulations to verify our conclusions. Simulation results show the correctness of our analysis on phase ambiguity and the effectiveness of our method.
(4) We have designed a two-stage algorithm of multiple beam generation, and conduct simulations on 19-element and 127-element CBC systems for Bi-beam generation. The results show the effectiveness of our method, which is expected to be applied for obtaining larger scale beams.

## 2. Principle

*2.1 Discussion on phase ambiguity*

The diagram of concept experiment in CBC system is shown in Fig. 1. We use a mask with an array of hard apertures to filter the broadened laser beam, so that we can obtain the equivalent beam splitting laser emitted by a specific array. The spatial light modulator (SLM1) is able to control the phases of each subbeam, which is employed to mimicking unknown piston aberrations. We use the second phase modulator (SLM2) for phase correction. The corrected subbeams are focused by a transform lens. The focused subbeam is sent to a 10× micro-objective (MO) and detected by a CCD camera to obtain the far-field image for phase-prediction algorithms.

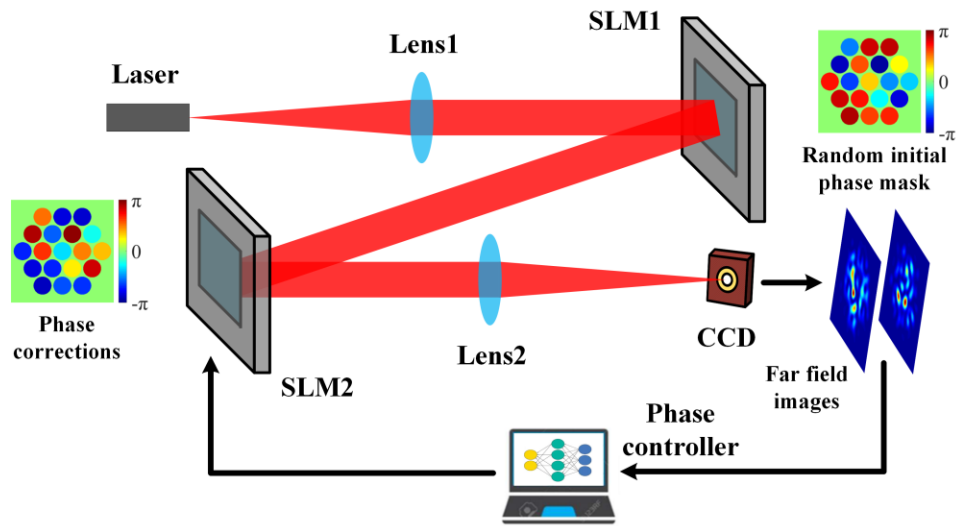

Fig. 1. The structure of N-elements CBC system. (N=19 in this diagram).

On the emission plane, the subapertures are usually arranged in the shape of regular hexagon. This is because the regular hexagon is the most compact arrangement scheme, which is beneficial to reduce sidelobe energy and obtain high quality combined beams [28-33]. However, regular hexagonal arrangement belongs to centrosymmetric arrangement, which will cause potential phase ambiguity.

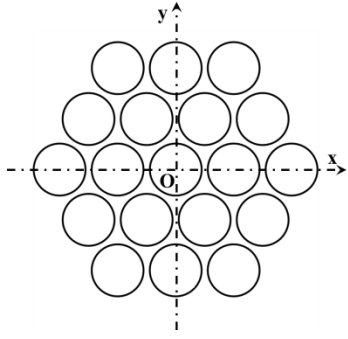

Fig. 2. Schematic diagram of emission plane coordinate system.

Without loss of generality, we construct a Cartesian coordinate system with the center of the emission plane as the origin, as shown in Fig. 2. The near-field complex amplitude of the *n-th* subbeam can be expressed as

$$U_{near_n}(x,y,0) = A_0 \cdot \exp\left[-\frac{(x-x_n)^2 + (y-y_n)^2}{\omega_0^2}\right] \cdot circ\left(\frac{(x-x_n)^2 + (y-y_n)^2}{(d/2)^2}\right) \quad (1)$$

where $A_0$ is the amplitude, $\omega_0$ is the waist radius (when the subbeams are plane waves, $\omega_0$ can be regarded as infinity), *d* is the diameter of subapertures, and ($x_n$, $y_n$, 0) is the coordinate of the *n-th* subaperture's center in the above coordinate system. $circ(r) = \begin{cases} 1, & r \leq 1 \\ 0, & \text{otherwise} \end{cases}$. The near-field complex amplitude after piston phase modulation could be expressed as

$$E_{near_n} = A_0 \cdot \exp\left[-\frac{(x-x_n)^2 + (y-y_n)^2}{\omega_0^2}\right] \cdot \exp[i \cdot \psi_n] \cdot circ\left(\frac{(x-x_n)^2 + (y-y_n)^2}{(d/2)^2}\right). \quad (2)$$

The index of the subaperture at the center is 0, the corresponding piston aberration is $\psi_0$ ( the value range of all phases in this paper is [-π, π]), and the corresponding center coordinate is $\begin{cases} x_0 = 0 \\ y_0 = 0 \end{cases}$. We find that when the subapertures are in centrosymmetric arrangement, for the *n-th* subaperture (n≠0), there always exists another subaperture which is centrosymmetric with it respect to the subaperture0. We denote the index of this subaperture as -n, the corresponding piston aberration is $\psi_{-n}$, and the central coordinate is ($x_{-n}$, $y_{-n}$, 0). It is obvious that $x_{-n}=-x_n$ and $y_{-n}=-y_n$. The overall near-field complex amplitude modulated by the piston phase is equal to the sum of each subapertures, which can be expressed as

$$E_{near} = \sum A_0 \cdot \exp\left[-\frac{(x-x_n)^2+(y-y_n)^2}{\omega_0^2}\right] \cdot \exp[i \cdot \psi_n] \cdot circ\left(\frac{(x-x_n)^2+(y-y_n)^2}{(d/2)^2}\right), \quad (3)$$

where n=0, ±1, ±2, ···, ±(N-1)/2. The near-field complex amplitude $E_{near}$ is transformed to the far-field complex amplitude $E_{far}$ at the focal plane by the lens. Then the relationship between $E_{far}$ and $E_{near}$ can be expressed by Fraunhofer diffraction formula as follows [34, 35].

$$\begin{aligned}
E_{far} &= \iint E_{near} \cdot \exp[-i \cdot 2\pi \cdot (ux+vy)] dxdy \\
&= \iint \sum A_0 \cdot \exp\left[-\frac{(x-x_n)^2+(y-y_n)^2}{\omega_0^2}\right] \cdot \exp[i \cdot \psi_n] circ\left(\frac{(x-x_n)^2+(y-y_n)^2}{(d/2)^2}\right) \\
&\quad \exp[-i \cdot 2\pi \cdot (ux+vy)] dxdy \\
&= \sum_n \iint_{circ_n} A_0 \cdot \exp\left[-\frac{(x-x_n)^2+(y-y_n)^2}{\omega_0^2}\right] \cdot \exp[i \cdot \psi_n] \cdot \exp[-i \cdot 2\pi \cdot (ux+vy)] dxdy
\end{aligned}$$

(4)

where $u = \frac{x_{far}}{\lambda f_{lens}}$ and $v = \frac{y_{far}}{\lambda f_{lens}}$ (λ is the beam wavelength, and $f_{lens}$ is the focal length of the transform lens). If we make the variable substitution that $x'=x-x_n$ and $y'=y-y_n$, the far-field complex amplitude of the $n$-$th$ subbeam can be rewritten as

$$\begin{aligned}
E_{far_n} &= \iint_{cir_0} A_0 \cdot \exp\left[-\frac{(x')^2+(y')^2}{\omega_0^2}+i \cdot \psi_n - i \cdot 2\pi \cdot [u(x'+x_n)+v(y'+y_n)]\right] dx'dy' \\
&= \iint_{cir_0} A_0 \cdot \exp\left[-\frac{(x')^2}{\omega_0^2}-i \cdot 2\pi \cdot u \cdot x'-\frac{(y')^2}{\omega_0^2}-i \cdot 2\pi \cdot v \cdot y'-i \cdot 2\pi \cdot x_n \cdot u - i \cdot 2\pi \cdot y_n \cdot v + i \cdot \psi_n\right] dx'dy'
\end{aligned}$$

.(5)

We separate the variables independent of $x'$ and $y'$, then move them out of the integral symbol. Equation (5) is transformed into

$$E_{far_n} = A_0 \cdot \exp[-i \cdot 2\pi \cdot x_n \cdot u - i \cdot 2\pi \cdot y_n \cdot v + i \cdot \psi_n] \cdot$$
$$\iint_{cir_0} \exp\left[-\frac{(x')^2}{\omega_0^2}-i \cdot 2\pi \cdot u \cdot x'-\frac{(y')^2}{\omega_0^2}-i \cdot 2\pi \cdot v \cdot y'\right] dx'dy' \quad .(6)$$

We denote $A(u,v) = A_0 \cdot \iint_{cir_0} \exp\left[-\frac{(x')^2}{\omega_0^2}-i \cdot 2\pi \cdot u \cdot x'-\frac{(y')^2}{\omega_0^2}-i \cdot 2\pi \cdot v \cdot y'\right] dx'dy'$ (the value

of $A(u,v)$ is independent of n) and substitute equation (6) into equation (4), then we obtain the superposition of each subbeam's far-field complex amplitude

$$E_{far}(u,v,\Psi) = \sum A(u,v) \cdot \exp[-i \cdot 2\pi \cdot x_n \cdot u - i \cdot 2\pi \cdot y_n \cdot v + i \cdot \psi_n]$$
$$= A(u,v) \cdot \sum \exp[-i \cdot 2\pi \cdot x_n \cdot u - i \cdot 2\pi \cdot y_n \cdot v + i \cdot \psi_n] \quad , \tag{7}$$

where $\Psi = \{\psi_n\}$ (n=0,±1,±2,⋯, ±(N−1)/2) denotes the set of the *n-th* subbeam's piston aberration.

The far-field intensity distribution $I_{far}$ could be calculated from the square of the far-field complex amplitude's modulus, which can be expressed as

$$I_{far} = \|E_{far}\|^2 = \|A(u,v)\|^2 \cdot \left\|\sum \exp[-i \cdot 2\pi \cdot x_n \cdot u - i \cdot 2\pi \cdot y_n \cdot v + i \cdot \psi_n]\right\|^2. \tag{8}$$

The far-field intensity distribution determines the shape of the far-field image on the charge coupled device (CCD) camera. Here, we only consider the influence of the piston aberration on the far-field image, so we separate $I'_{far}$ ($I_{far} = \|A(u,v)\|^2 \cdot I'_{far}$), which is related to $\Psi = \{\psi_n\}$.

$$\begin{aligned}
I'_{far}(\Psi) &= \left\|\sum \exp[-i \cdot 2\pi \cdot x_n \cdot u - i \cdot 2\pi \cdot y_n \cdot v + i \cdot \psi_n]\right\|^2 \\
&= \left\|\sum \cos(-2\pi \cdot x_n \cdot u - 2\pi \cdot y_n \cdot v + \psi_n) + i \cdot \sin(-2\pi \cdot x_n \cdot u - 2\pi \cdot y_n \cdot v + \psi_n)\right\|^2 \\
&= \left[\sum \cos(-2\pi \cdot x_n \cdot u - 2\pi \cdot y_n \cdot v + \psi_n)\right]^2 + \left[\sum \sin(-2\pi \cdot x_n \cdot u - 2\pi \cdot y_n \cdot v + \psi_n)\right]^2 \\
&= \sum_i \sum_j \cos(-2\pi \cdot x_i \cdot u - 2\pi \cdot y_i \cdot v + \psi_i) \cdot \cos(-2\pi \cdot x_j \cdot u - 2\pi \cdot y_j \cdot v + \psi_j) + \\
&\quad \sum_i \sum_j \sin(-2\pi \cdot x_i \cdot u - 2\pi \cdot y_i \cdot v + \psi_i) \cdot \sin(-2\pi \cdot x_j \cdot u - 2\pi \cdot y_j \cdot v + \psi_j) \\
&= \sum_i \sum_j \cos[-2\pi \cdot (x_i - x_j) \cdot u - 2\pi \cdot (y_i - y_j) \cdot v + (\psi_i - \psi_j)]
\end{aligned} \tag{9}$$

where i,j=0,±1,±2,⋯, ±(N−1)/2. Without loss of generality, we have $\varphi_i = \psi_i - \psi_0$, then a set of N pistons $\Psi = \{\psi_n\}$ (n=0,±1,±2,…, ±(N−1)/2) can be expressed by a set of N-1 relative pistons $\Phi = \{\varphi_n\}$ (n=±1,±2,…, ±(N−1)/2). This corresponds to phase redundancy (More details are in Appendix 1.). After substituting the relative pistons $\Phi = \{\varphi_n\}$ and $\begin{cases} x_0 = 0 \\ y_0 = 0 \end{cases}$ into equation (9), we get

$$I'_{far}(\Phi) = N + \sum_i \cos(-2\pi \cdot x_i \cdot u - 2\pi \cdot y_i \cdot v + \varphi_i) + \\
\sum_i \sum_j \cos[-2\pi \cdot (x_i - x_j) \cdot u - 2\pi \cdot (y_i - y_j) \cdot v + (\varphi_i - \varphi_j)], \tag{10}$$

where i, j= ± 1, ± 2, ⋯, ± (N − 1)/2. We define $r_i(u,v) = -2\pi \cdot x_i \cdot u - 2\pi \cdot y_i \cdot v$ and $r_{i,j} = r_i - r_j = -2\pi \cdot (x_i - x_j) \cdot u - 2\pi \cdot (y_i - y_j) \cdot v$. It is obvious that $-r_i = r_{-i}$ and $-r_{i,j} = r_{-i,-j}$. Then we define $B_0(\Phi) = \sum_i \cos(-2\pi \cdot x_i \cdot u - 2\pi \cdot y_i \cdot v + \varphi_i)$ and $B_j(\Phi) = \sum_i \cos[-2\pi \cdot (x_i - x_j) \cdot u$

$-2\pi \cdot (y_i - y_j) \cdot v + (\varphi_i - \varphi_j)]$ (when j≠0). By substituting them into equation (10), we obtain

$$I'_{far}(\Phi) = N + B_0(\Phi) + \sum_j B_j(\Phi), \quad (11)$$

$$B_0(\Phi) = \begin{bmatrix} \cos(r_1) \\ \cos(r_2) \\ ... \\ \cos(r_{(N-1)/2}) \end{bmatrix}^T \cdot \begin{bmatrix} \cos(\varphi_1) \\ \cos(\varphi_2) \\ ... \\ \cos(\varphi_{(N-1)/2}) \end{bmatrix} + \begin{bmatrix} \cos(r_{-1}) \\ \cos(r_{-2}) \\ ... \\ \cos(r_{-(N-1)/2}) \end{bmatrix}^T \cdot \begin{bmatrix} \cos(\varphi_{-1}) \\ \cos(\varphi_{-2}) \\ ... \\ \cos(\varphi_{-(N-1)/2}) \end{bmatrix}$$

$$- \begin{bmatrix} \sin(r_1) \\ \sin(r_2) \\ ... \\ \sin(r_{(N-1)/2}) \end{bmatrix}^T \cdot \begin{bmatrix} \sin(\varphi_1) \\ \sin(\varphi_2) \\ ... \\ \sin(\varphi_{(N-1)/2}) \end{bmatrix} - \begin{bmatrix} \sin(r_{-1}) \\ \sin(r_{-2}) \\ ... \\ \sin(r_{-(N-1)/2}) \end{bmatrix}^T \cdot \begin{bmatrix} \sin(\varphi_{-1}) \\ \sin(\varphi_{-2}) \\ ... \\ \sin(\varphi_{-(N-1)/2}) \end{bmatrix}$$

$$= \begin{bmatrix} \cos(r_1) \\ \cos(r_2) \\ ... \\ \cos(r_{(N-1)/2}) \end{bmatrix}^T \cdot \begin{bmatrix} \cos(\varphi_1) \\ \cos(\varphi_2) \\ ... \\ \cos(\varphi_{(N-1)/2}) \end{bmatrix} + \begin{bmatrix} \cos(-r_1) \\ \cos(-r_2) \\ ... \\ \cos(-r_{(N-1)/2}) \end{bmatrix}^T \cdot \begin{bmatrix} \cos(\varphi_{-1}) \\ \cos(\varphi_{-2}) \\ ... \\ \cos(\varphi_{-(N-1)/2}) \end{bmatrix}$$

$$- \begin{bmatrix} \sin(r_1) \\ \sin(r_2) \\ ... \\ \sin(r_{(N-1)/2}) \end{bmatrix}^T \cdot \begin{bmatrix} \sin(\varphi_1) \\ \sin(\varphi_2) \\ ... \\ \sin(\varphi_{(N-1)/2}) \end{bmatrix} - \begin{bmatrix} \sin(-r_1) \\ \sin(-r_2) \\ ... \\ \sin(-r_{(N-1)/2}) \end{bmatrix}^T \cdot \begin{bmatrix} \sin(\varphi_{-1}) \\ \sin(\varphi_{-2}) \\ ... \\ \sin(\varphi_{-(N-1)/2}) \end{bmatrix}$$

$$= \begin{bmatrix} \cos(r_1) \\ \cos(r_2) \\ ... \\ \cos(r_{(N-1)/2}) \end{bmatrix}^T \cdot \begin{bmatrix} \cos(\varphi_1)+\cos(\varphi_{-1}) \\ \cos(\varphi_2)+\cos(\varphi_{-2}) \\ ... \\ \cos(\varphi_{(N-1)/2})+\cos(\varphi_{-(N-1)/2}) \end{bmatrix} - \begin{bmatrix} \sin(r_1) \\ \sin(r_2) \\ ... \\ \sin(r_{(N-1)/2}) \end{bmatrix}^T \cdot \begin{bmatrix} \sin(\varphi_1)-\sin(\varphi_{-1}) \\ \sin(\varphi_2)-\sin(\varphi_{-2}) \\ ... \\ \sin(\varphi_{(N-1)/2})-\sin(\varphi_{-(N-1)/2}) \end{bmatrix}$$

,(12)

$$B_j(\Phi) = \begin{bmatrix} \cos(r_{1,j}) \\ \cos(r_{2,j}) \\ ... \\ \cos(r_{(N-1)/2,j}) \end{bmatrix}^T \cdot \begin{bmatrix} \cos(\varphi_1 - \varphi_j) \\ \cos(\varphi_2 - \varphi_j) \\ ... \\ \cos(\varphi_{(N-1)/2} - \varphi_j) \end{bmatrix} + \begin{bmatrix} \cos(r_{-1,j}) \\ \cos(r_{-2,j}) \\ ... \\ \cos(r_{-(N-1)/2,j}) \end{bmatrix}^T \cdot \begin{bmatrix} \cos(\varphi_{-1} - \varphi_j) \\ \cos(\varphi_{-2} - \varphi_j) \\ ... \\ \cos(\varphi_{-(N-1)/2} - \varphi_j) \end{bmatrix}$$
$$- \begin{bmatrix} \sin(r_{1,j}) \\ \sin(r_{2,j}) \\ ... \\ \sin(r_{(N-1)/2,j}) \end{bmatrix}^T \cdot \begin{bmatrix} \sin(\varphi_1 - \varphi_j) \\ \sin(\varphi_2 - \varphi_j) \\ ... \\ \sin(\varphi_{(N-1)/2} - \varphi_j) \end{bmatrix} - \begin{bmatrix} \sin(r_{-1,j}) \\ \sin(r_{-2,j}) \\ ... \\ \sin(r_{-(N-1)/2,j}) \end{bmatrix}^T \cdot \begin{bmatrix} \sin(\varphi_{-1} - \varphi_j) \\ \sin(\varphi_{-2} - \varphi_j) \\ ... \\ \sin(\varphi_{-(N-1)/2} - \varphi_j) \end{bmatrix}$$
,(13)

$$B_{-j}(\Phi) = \begin{bmatrix} \cos(r_{1,-j}) \\ \cos(r_{2,-j}) \\ ... \\ \cos(r_{(N-1)/2,-j}) \end{bmatrix}^T \cdot \begin{bmatrix} \cos(\varphi_1 - \varphi_{-j}) \\ \cos(\varphi_2 - \varphi_{-j}) \\ ... \\ \cos(\varphi_{(N-1)/2} - \varphi_{-j}) \end{bmatrix} + \begin{bmatrix} \cos(r_{-1,-j}) \\ \cos(r_{-2,-j}) \\ ... \\ \cos(r_{-(N-1)/2,-j}) \end{bmatrix}^T \cdot \begin{bmatrix} \cos(\varphi_{-1} - \varphi_{-j}) \\ \cos(\varphi_{-2} - \varphi_{-j}) \\ ... \\ \cos(\varphi_{-(N-1)/2} - \varphi_{-j}) \end{bmatrix}$$
$$- \begin{bmatrix} \sin(r_{1,-j}) \\ \sin(r_{2,-j}) \\ ... \\ \sin(r_{(N-1)/2,-j}) \end{bmatrix}^T \cdot \begin{bmatrix} \sin(\varphi_1 - \varphi_{-j}) \\ \sin(\varphi_2 - \varphi_{-j}) \\ ... \\ \sin(\varphi_{(N-1)/2} - \varphi_{-j}) \end{bmatrix} - \begin{bmatrix} \sin(r_{-1,-j}) \\ \sin(r_{-2,-j}) \\ ... \\ \sin(r_{-(N-1)/2,-j}) \end{bmatrix}^T \cdot \begin{bmatrix} \sin(\varphi_{-1} - \varphi_{-j}) \\ \sin(\varphi_{-2} - \varphi_{-j}) \\ ... \\ \sin(\varphi_{-(N-1)/2} - \varphi_{-j}) \end{bmatrix}$$
, (14)
$$= \begin{bmatrix} \cos(-r_{1,j}) \\ \cos(-r_{2,j}) \\ ... \\ \cos(-r_{(N-1)/2,j}) \end{bmatrix}^T \cdot \begin{bmatrix} \cos(\varphi_1 - \varphi_{-j}) \\ \cos(\varphi_2 - \varphi_{-j}) \\ ... \\ \cos(\varphi_{(N-1)/2} - \varphi_{-j}) \end{bmatrix} + \begin{bmatrix} \cos(-r_{-1,j}) \\ \cos(-r_{-2,j}) \\ ... \\ \cos(-r_{-(N-1)/2,j}) \end{bmatrix}^T \cdot \begin{bmatrix} \cos(\varphi_{-1} - \varphi_{-j}) \\ \cos(\varphi_{-2} - \varphi_{-j}) \\ ... \\ \cos(\varphi_{-(N-1)/2} - \varphi_{-j}) \end{bmatrix}$$
$$- \begin{bmatrix} \sin(-r_{1,j}) \\ \sin(-r_{2,j}) \\ ... \\ \sin(-r_{(N-1)/2,j}) \end{bmatrix}^T \cdot \begin{bmatrix} \sin(\varphi_1 - \varphi_{-j}) \\ \sin(\varphi_2 - \varphi_{-j}) \\ ... \\ \sin(\varphi_{(N-1)/2} - \varphi_{-j}) \end{bmatrix} - \begin{bmatrix} \sin(-r_{-1,j}) \\ \sin(-r_{-2,j}) \\ ... \\ \sin(-r_{-(N-1)/2,j}) \end{bmatrix}^T \cdot \begin{bmatrix} \sin(\varphi_{-1} - \varphi_{-j}) \\ \sin(\varphi_{-2} - \varphi_{-j}) \\ ... \\ \sin(\varphi_{-(N-1)/2} - \varphi_{-j}) \end{bmatrix}$$

We find that $\cos(r_i)$ is linearly correlated with $\cos(r_j)$ only if i=j or i=-j. At the same time, $\cos(r_{i1,j1})$ is linearly related to $\cos(r_{i2,j2})$, only if $|r_{i1,j1}| = |r_{i2,j2}|$. By analyzing the variables related to u and v in $B_j(\Phi)$, we find that when i≠j and i≠-j, $B_i(\Phi)$ is not linearly correlated with $B_j(\Phi)$. Therefore, if $\Phi = \{\varphi_n\}$ and $\Theta = \{\theta_n\}$ generate the same far-field image, that is

$I'_{far}(\Phi) = I'_{far}(\Theta)$, there must be

$$B_0(\Phi) = B_0(\Theta), \quad (15)$$

$$B_i(\Phi) + B_{-i}(\Phi) = B_i(\Theta) + B_{-i}(\Theta). \quad (16)$$

Considering the expression of equation (12), if equation (15) holds, the coefficients of the corresponding components are equal, which could be expressed as

$$\begin{cases} \cos(\varphi_i) + \cos(\varphi_{-i}) = \cos(\theta_i) + \cos(\theta_{-i}) \\ \sin(\varphi_i) - \sin(\varphi_{-i}) = \sin(\theta_i) - \sin(\theta_{-i}) \end{cases}. \quad (17)$$

$$\Rightarrow (\cos(\varphi_i) + \cos(\varphi_{-i}) - \cos(\theta_i))^2 + (\sin(\varphi_i) - \sin(\varphi_{-i}) - \sin(\theta_i))^2 = (\cos(\theta_{-i}))^2 + (-\sin(\theta_{-i}))^2 = 1$$

$$3 + 2\cos(\varphi_i) \cdot \cos(\varphi_{-i}) - 2[\cos(\varphi_i) + \cos(\varphi_{-i})] \cdot \cos(\theta_i)$$

$$-2\sin(\varphi_i) \cdot \sin(\varphi_{-i}) - 2[\sin(\varphi_i) - \sin(\varphi_{-i})] \cdot \sin(\theta_i) = 1$$

$$2 + 2\cos(\varphi_i + \varphi_{-i}) = 2[\cos(\varphi_i) + \cos(\varphi_{-i})] \cdot \cos(\theta_i) + 2[\sin(\varphi_i) - \sin(\varphi_{-i})] \cdot \sin(\theta_i)$$

$$4\cos^2 \frac{\varphi_i + \varphi_{-i}}{2} = 2\left[2\cos(\frac{\varphi_i + \varphi_{-i}}{2}) \cdot \cos(\frac{\varphi_i - \varphi_{-i}}{2})\right] \cdot \cos(\theta_i) + 2\left[2\cos(\frac{\varphi_i + \varphi_{-i}}{2}) \cdot \sin(\frac{\varphi_i - \varphi_{-i}}{2})\right] \cdot \sin(\theta_i)$$

$$4\cos^2 \frac{\varphi_i + \varphi_{-i}}{2} = 4\cos(\frac{\varphi_i + \varphi_{-i}}{2}) \cdot \cos(\frac{\varphi_i - \varphi_{-i}}{2} - \theta_i)$$

(18)

There are 3 solutions for equation (18), which are

$$\begin{cases} \varphi_i + \varphi_{-i} = \pi \\ \theta_i + \theta_{-i} = \pi \end{cases}, (19) \quad \begin{cases} \theta_i = \varphi_i \\ \theta_{-i} = \varphi_{-i} \end{cases}, (20) \quad \begin{cases} \theta_i = -\varphi_{-i} \\ \theta_{-i} = -\varphi_i \end{cases}. (21)$$

We denote the set of $\theta_i$, $\theta_{-i}$ which satisfies equation (19) as $E_1$, the set of $\theta_i$, $\theta_{-i}$ which satisfies equation (20) as $E_2$, the set of $\theta_i$, $\theta_{-i}$ which satisfies equation (21) as $E_3$. Therefore, if $\Phi = \{\varphi_n\}$ and $\Theta = \{\theta_n\}$ generate the same far-field image, $\theta_n \in E_1$ or $\theta_n \in E_2$ or $\theta_n \in E_3$. When $\theta_i, \theta_j \in E_1$, we substitute them into equation (15) and get

$$\begin{cases} \cos(\varphi_i - \varphi_j) + \cos(\varphi_{-i} - \varphi_{-j}) = \cos(\theta_i - \theta_j) + \cos(\theta_{-i} - \theta_{-j}) \\ \sin(\varphi_i - \varphi_j) - \sin(\varphi_{-i} - \varphi_{-j}) = \sin(\theta_i - \theta_j) - \sin(\theta_{-i} - \theta_{-j}) \end{cases}$$

$$\Rightarrow \begin{cases} \cos(\varphi_i - \varphi_j) + \cos((\pi - \varphi_i) - (\pi - \varphi_j)) = 2\cos(\varphi_i - \varphi_j) = 2\cos(\theta_i - \theta_j) \\ \sin(\varphi_i - \varphi_j) - \sin((\pi - \varphi_i) - (\pi - \varphi_j)) = 2\sin(\varphi_i - \varphi_j) = 2\sin(\theta_i - \theta_j) \end{cases}, (22)$$

$$\Rightarrow \theta_i = \varphi_i + \Delta, (\Delta = 0 \text{ or } \Delta = \pi)$$

We find that $\theta_i \in E_2$ when $\Delta = 0$ and $\theta_i \in E_3$ when $\Delta = \pi$. Therefore, the above three solutions can be combined into two. That is, if $\Phi = \{\varphi_n\}$ and $\Theta = \{\theta_n\}$ generate the same far-field image, $\theta_n \in E_2$ or $\theta_n$

$\in E_3$. It is obvious that equation (16) holds when $\theta_i$, $\theta_j \in E_2$. Similarly, equation (16) holds when $\theta_i$, $\theta_j \in E_3$.

Obviously, $I'_{far}(\Phi) = I'_{far}(\Theta)$ has a special solution (*special solution 1*) that all elements in $\Theta = \{\theta_n\}$ satisfy equation (20). Another special solution (*special solution 2*) is that all elements in $\Theta = \{\theta_n\}$ satisfy equation (21).

Next, we consider whether there exists other solutions $\Theta = \{\theta_n\}$, in which some elements satisfy equation (20) and other elements satisfy equation (21). When $\theta_i \in E_2$ and $\theta_j \in E_3$, and equation (16) holds, we get

$$\begin{cases} \cos(\varphi_i - \varphi_j) + \cos(\varphi_{-i} - \varphi_{-j}) = \cos(\theta_i - \theta_j) + \cos(\theta_{-i} - \theta_{-j}) \\ \sin(\varphi_i - \varphi_j) - \sin(\varphi_{-i} - \varphi_{-j}) = \sin(\theta_i - \theta_j) - \sin(\theta_{-i} - \theta_{-j}) \end{cases}$$
$$\Rightarrow \begin{cases} \cos(\varphi_i - \varphi_j) + \cos(\varphi_{-i} - \varphi_{-j}) = \cos(\varphi_i + \varphi_{-j}) + \cos(\varphi_{-i} + \varphi_j) \\ \sin(\varphi_i - \varphi_j) - \sin(\varphi_{-i} - \varphi_{-j}) = \sin(\varphi_i + \varphi_{-j}) - \sin(\varphi_{-i} + \varphi_j) \end{cases} \tag{23}$$

Similar to equations (19), (20) and (21), equation (23) also has three solutions

$$\begin{cases} \varphi_i - \varphi_j + \varphi_{-i} - \varphi_{-j} = \pi \\ \varphi_i + \varphi_{-j} + \varphi_{-i} + \varphi_j = \pi \end{cases} \Rightarrow \varphi_j + \varphi_{-j} = 0 \Rightarrow \begin{cases} \theta_j = \varphi_j \\ \theta_{-j} = \varphi_{-j} \end{cases}, \tag{24}$$

$$\begin{cases} \varphi_i - \varphi_j = \varphi_i + \varphi_{-j} \\ \varphi_{-i} - \varphi_{-j} = \varphi_{-i} + \varphi_j \end{cases} \Rightarrow \varphi_j + \varphi_{-j} = 0 \Rightarrow \begin{cases} \theta_j = \varphi_j \\ \theta_{-j} = \varphi_{-j} \end{cases}, \tag{25}$$

$$\begin{cases} -\varphi_{-i} + \varphi_{-j} = \varphi_i + \varphi_{-j} \\ -\varphi_i + \varphi_j = \varphi_{-i} + \varphi_j \end{cases} \Rightarrow \varphi_i + \varphi_{-i} = 0 \Rightarrow \begin{cases} \theta_i = -\varphi_{-i} \\ \theta_{-i} = -\varphi_{-i} \end{cases}, \tag{26}$$

When equation (24) holds, $\theta_j \in E_2$, which corresponds to the case of *special solution 1*.
When equation (25) holds, $\theta_j \in E_2$, which corresponds to the case of *special solution 1*.
When equation (26) holds, $\theta_i \in E_3$, which corresponds to the case of *special solution 2*.

Similarly, when $\theta_i \in E_3$ and $\theta_j \in E_2$, and equation (16) holds, the solution belongs to *special solution 1* or *special solution 2* as well. Therefore, *special solution 1* and *special solution 2* constitute all solutions of $I'_{far}(\Phi) = I'_{far}(\Theta)$. That is, when $\Phi = \{\varphi_n\}$ and $\Theta = \{\theta_n\}$ generate the same far-field image, there exists $\theta_n = \varphi_n$ (*solution 1*) or $\theta_n = -\varphi_{-n}$ (*solution 2*) for n=±1,±2,…, ±(N−1)/2.

### 2.2 Solution to phase ambiguity

According to the analysis in Section 2.1, *solution 1* corresponds to the original piston aberration (denoted as $\Phi$), and *solution 2* corresponds to the phase obtained by rotating the original piston aberration 180 degrees and conjugating it (denoted as $\overline{\Phi}$). Therefore, $\Phi$ and its rotationally conjugate piston aberration $\overline{\Phi}$ will generate the same far field. This multi solution problem is the phase ambiguity. In CBC, it is necessary to distinguish these two situations so that we could obtain the accurate information of piston aberration. In practical application, in order to achieve this, additional optical device needs to be added to the system, which will increase the cost of building and maintaining the experimental platform. Especially when the system is highly integrated, it is difficult to add additional optical devices. We consider distinguishing these two piston aberrations by breaking the rotationally conjugate symmetry in principle. We declare that using non-centrosymmetric arrays is a choice to solve this problem, but as mentioned earlier, it will damage the quality of the combined beam. Our method is to introduce phase modulation $\Lambda = \{\gamma_n\}$. The modulated phase can be expressed as $\Phi_m = \{\varphi_n + \gamma_n\}$ and $\overline{\Phi}_m = \{\overline{\varphi}_n + \gamma_n\}$, where $\overline{\varphi}_n = -\varphi_{-n}$. If $\Phi_m$ and $\overline{\Phi}_m$ generate different far-field images, the following formula holds.

$$\varphi_n + \gamma_n \neq \overline{\varphi}_n + \gamma_n \tag{27}$$

$$-\varphi_{-n} - \gamma_{-n} \neq \overline{\varphi}_n + \gamma_n \tag{28}$$

In fact, if $\varphi_n \neq -\varphi_{-n}$, equation (27) always holds. If $\varphi_n = -\varphi_{-n}$ (that is $\Phi = \overline{\Phi}$), the multi solution problem no longer existed. Therefore, we only consider equation (28). When it holds, there should be

$$\gamma_n \neq -\gamma_{-n}. \tag{29}$$

Thus, if $\Lambda = \{\gamma_n\}$ satisfied equation (29), we would discriminate $\Phi$ and $\overline{\Phi}$ ($\Phi \neq \overline{\Phi}$) by applying modulation phase. We only need to add an additional step to overcome the phase ambiguity without adding other optical devices, so as to obtain the accurate information of piston aberration through a pair of far-field images.

## 3. Simulation and result analysis

In Section 2, we analyze all the solutions of piston aberration that will cause the phase ambiguity in CBC system with centrosymmetric distribution of subapertures, and provide a method to solve this problem in theory. In this section, we will verify our conclusions of theoretical derivation through simulation.

We conduct simulations based on the CBC system shown in Fig. 1, and the parameters are A=1, $\omega_0$=11 mm, λ=1064 nm, and $f_{lens} = 2$ m. We first randomly generate 50 groups of piston aberration as *Target Phase*. Each group of *Target Phase* will generate its corresponding far-field image (*Target Far-field*, TFF). For each TFF, we first randomly initialize the value of *Same Far-field Phase* (*SFF Phase*), then we generate its far-field image (*Same Far-field*, *SFF*). We calculate the mean square error (MSE) loss between normalized *TFF* and *SFF*, and use this loss to optimize *SFF Phase*. When the loss is less than 1e−5, we stopped the iteration. At this time, it is considered that *Target Phase* and *SFF Phase* generated approximately same far-fields. The process is shown in Fig. 3

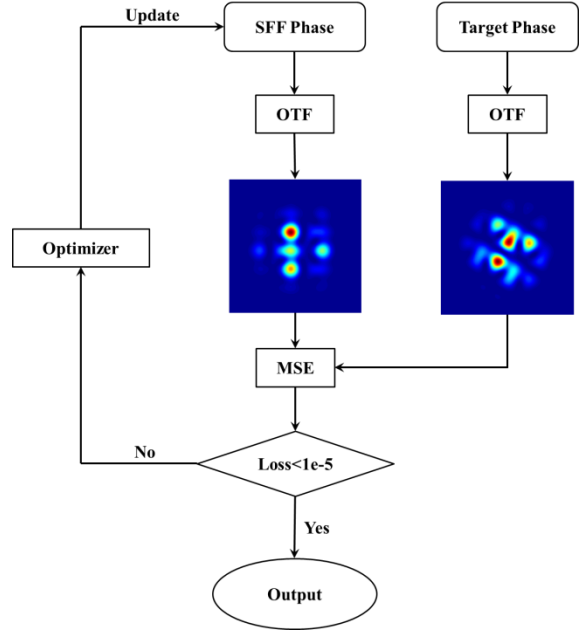

Fig. 3. The process diagram of how to find the piston aberration which generates the samiliar far-field to Target Phase.

For each *Target Phase*, we randomly initialize the value of *Same Far-field Phase* (*SFF Phase*) for 200 times, which means we will find 200 *SFF Phases* that has similar far-fields with *TFF*. If our theoretical derivation is correct, the solutions are distributed in two intervals. One is *Target Phase* itself, and the other is *Target Phase*'s rotationally conjugate piston aberration. If there remains other intervals, it indicates that other solutions exist. If there exists only one interval, it means that the rotationally conjugate piston aberration $\overline{\Phi}$ will not generate the same far-field image as the original one $\Phi$ (Another possibility is that $\Phi = \overline{\Phi}$, but we avoid generating this kind of *Target Phases*.).

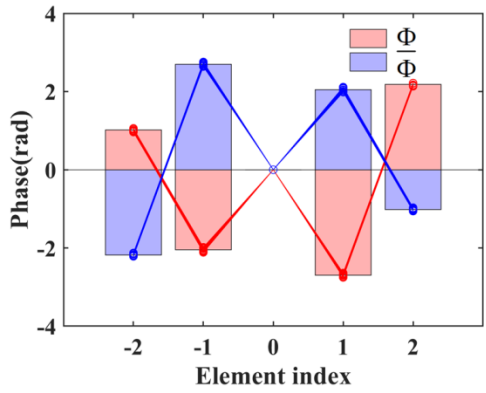
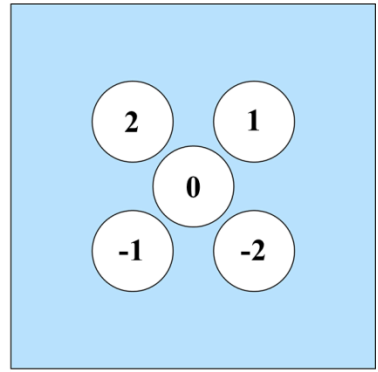

(a) (b)

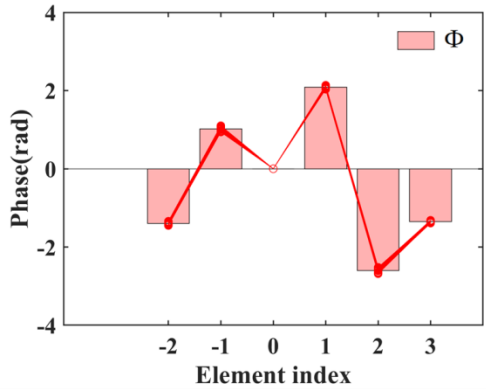
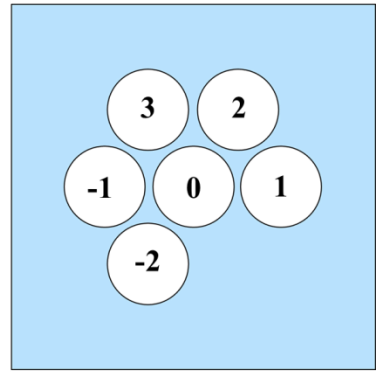

(c) (d)

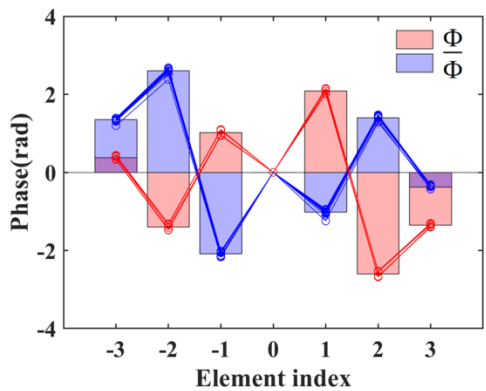
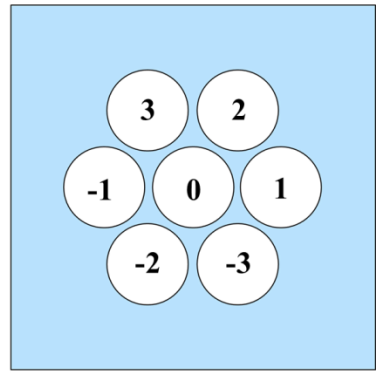

(e) (f)

Fig. 4. If we make *SFF Phase* generate the samiliar far field as *Target Phase*, the solutions are distributed in two intervals in centrosymmetric arrays (5-element in (a), (b), and 7-element in (e), (f)). However, there is a unique solution in non-centrosymmetric arrays (6-element in (c), (d)).

We conducted simulations on a 5-element, 6-element, and 7-element CBC system, and the results demonstrate that our conclusions are correct. We randomly selected a *Target Phase* for each system and draw the results of above simulation iterations as Fig. 4. In centrosymmetric-array CBC systems (5-element and 7-element), the solutions are distributed in two intervals. One is *Target Phase* itself, and the other is *Target Phase*'s rotationally conjugate piston aberration. Fig. 4 shows that $\overline{\Phi}$ will generate the same far-field image as $\Phi$ as well. In non-centrosymmetric arrays CBC system (6-element), the solutions only distribute near *Target Phase* itself.

Aiming at verifying whether our method can effectively solve the phase ambiguity problem, we designed a simulation process similar to that before. The difference is that we not only require *SFF Phase* and *Target Phase* to generate similar far-fields, but also require the modulated ones to generate similar far-fields. The process is shown in Fig. 5.

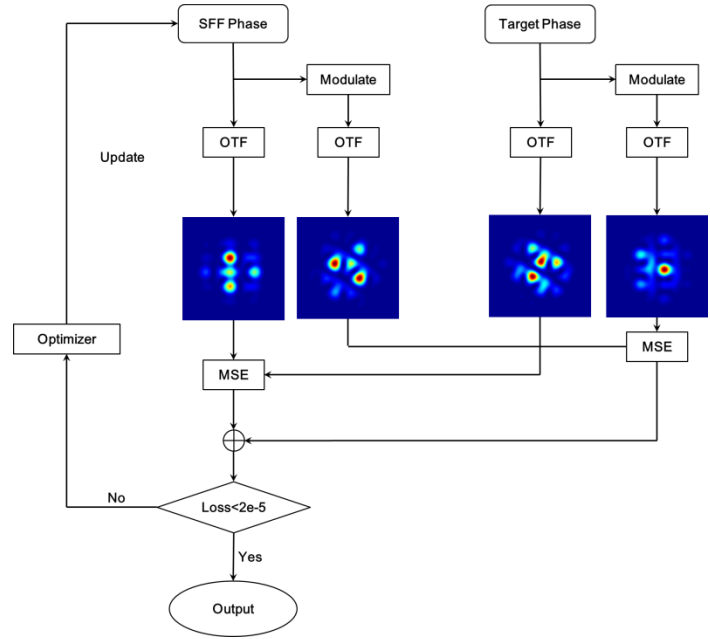

Fig. 5. The process diagram of how to find the piston aberration which generates the samiliar far-field to *Target Phase* with and without modulated.

If our method works, the solutions are only distributed near *Target Phase*. The results (shown in Fig. 6) of simulations on 5-element and 7-element are consistent with our conclusions, which indicates that our method is effective.

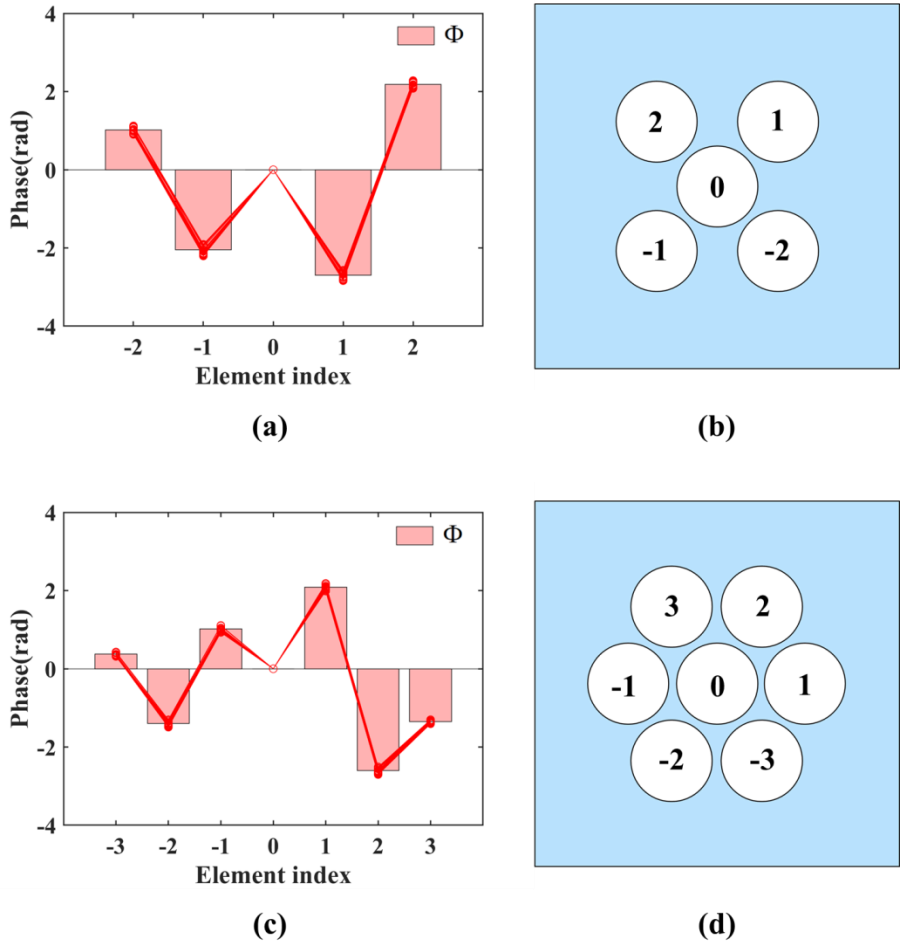

Fig. 6. The piston aberrations which generate the samiliar far-fields to *Target Phase* with and without modulated. All solutions are only distributed near *Target Phase*.

## 4. Multiple beam generation

Most CBC systems focus on generating a single high-power combined beam through realizing co-phase, lacking of the discussion on generating multiple beams. Actually, multiple beam generation plays an important role in Laser Transmission, Material Processing and so on. In order to generate multiple beams, we superimpose specific phases on the basis of reaching co-phase. We apply an optimization algorithm to obtain the corresponding phases. The core of the algorithm is to design an appropriate objective function. Some algorithms set the objective function as the beam intensity (which represents power) in specific areas. However, the problem is that the positions of multiple beams in the far field need to be determined in advance. For unknown arrays, these positions are usually unknown. Therefore, our algorithm needs to find the phases that generate multi-beams without a priori.

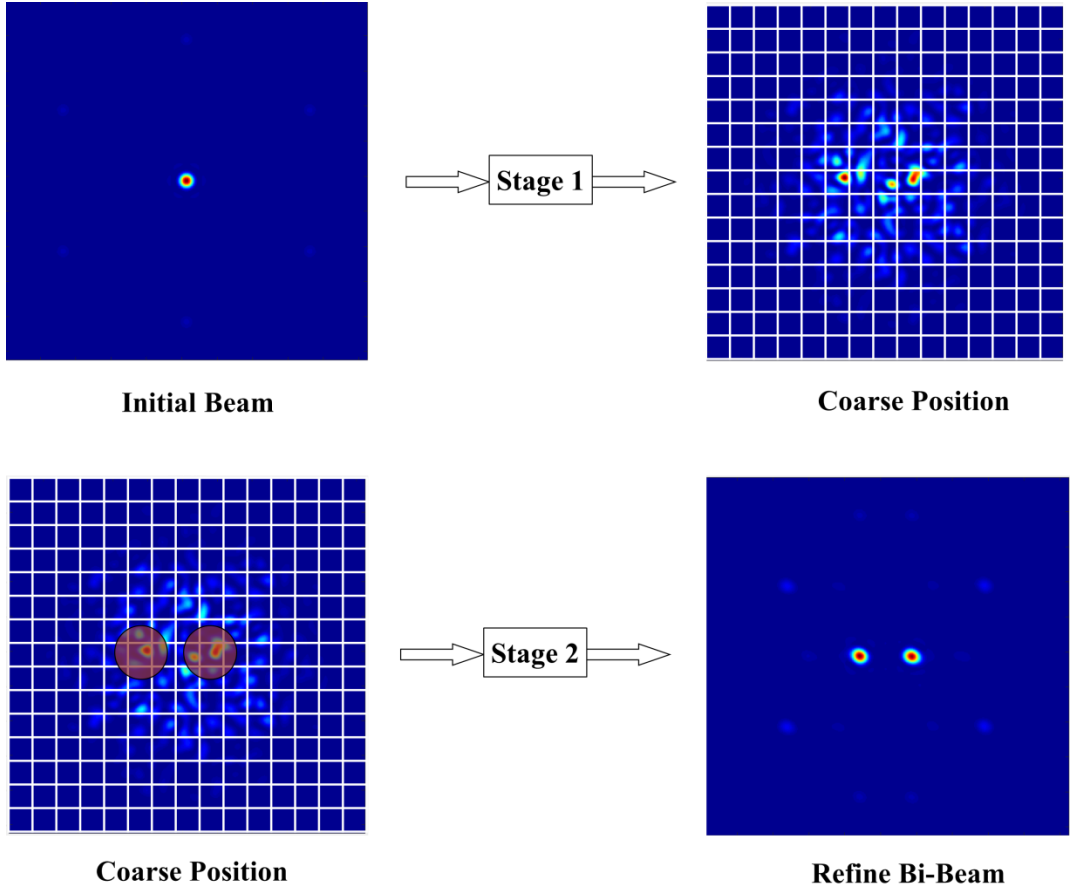

Fig. 7. The process of our two-stage Bi-beam generation algorithm.

We design a two-stage Bi-beam generation algorithm to verify our scheme. In the first stage, we divide the far field into N×N grids (N=15 in our work). Then, the grid with the greatest intensity is set as the position of primary beam, and the one with the second largest intensity is set as the position of secondary beam. Finally, we optimize the objective function. Our aim is to maximize the power of the primary beam and the secondary beam, and make them separate. Our objective function loss is as follows.

$$objective\ function\ loss1 = -I_{secondary} + I_{remaining} + loss_{separate} \qquad (30)$$

Where $I_{secondary}$ is the secondary beam's intensity, and the higher it is, the lower the loss is. This shows that one of the targets of Bi-beam shaping is to generate another beam with high energy besides the primary beam.

$I_{remaining}$ is the sum of the light intensities in the areas except the primary beam and the secondary beam. This term is introduced to suppress the generation of other high-energy beams (which prevents the generation of more than the target number of beams) and concentrate the energy in Bi-beam's areas.

$loss_{separate}$ is to measure the degree of separation between primary and secondary beams. If it is not

introduced, a strip spot may generate, and the primary beam and secondary beam are distributed in adjacent grids. At this time, although the energy of the secondary beam is high, it is not shaped into separated Bi-beam (which is a single beam essentially). Because the coordinates cannot transfer the gradient, we apply correlation calculation to measure the degree of beam separation. We take the primary beam as the center to generate the light intensity diffusion diagram, and generate the light intensity diffusion diagram of the secondary beam in the same way. Then we calculate the inner product between these two diagrams (the process is described in Fig. 8). When the primary beam and the secondary beam are closer together, $loss_{separate}$ value is high.

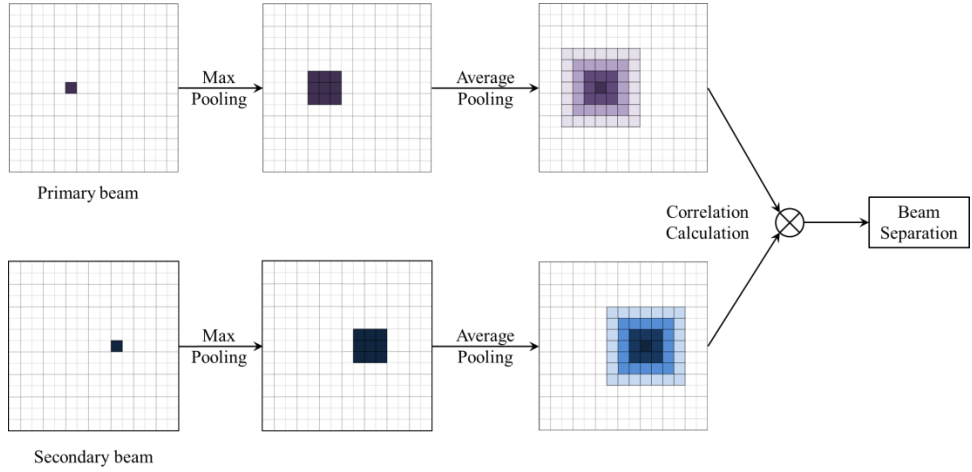

Fig. 8. We calculate $loss_{separate}$ through correlation calculation.

Through the algorithm in the first stage, we obtain the coarse position of Bi-beams in the far field (More discussions are in Appendix 2.). As shown in the Fig. 7, the coarse position is set as circular areas with grid as the center and 1.5 times grid side length as the radius. In the second stage, we apply refined optimization on this basis to obtain energy-concentrated Bi-beams. The objective function loss in this stage is defined as

$$objective\ function\ loss2 = -I_{primary}^{cir} - I_{secondary}^{cir} + k_0 \cdot (I_{primary}^{cir} - r_0 \cdot I_{secondary}^{cir})^2 . \qquad (31)$$

$-I_{primary}^{cir}$ and $-I_{secondary}^{cir}$ are designed to improve the energy in target circular areas.

$(I_{primary}^{cir} - r_0 \cdot I_{secondary}^{cir})^2$ is used to balance the energy of the primary beam and the secondary beam.

$r_0$ and $k_0$ were set to 1 when we generated balanced Bi-Beam.

The main parameters in our method are down-sampling scale factor, down-sampling mode, $r_0$ and

$k_0$. In this paper, we down-sample the far-field image into 15×15 grids with average-pooling. We simulated on 19-element and unit 127-element CBC system. The results are shown in the Fig. 9. It is noted that there is usually more than one phase list which may generate Bi-beam due to the rotational symmetry of arrays' arrangement.

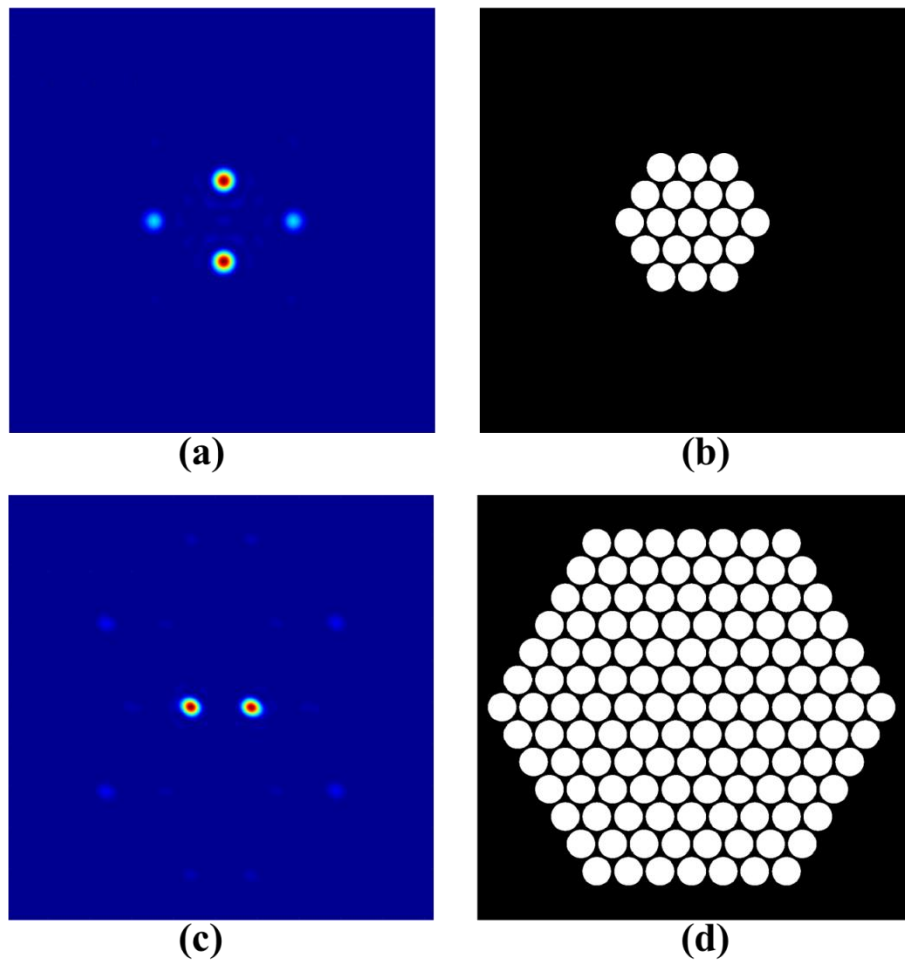

Fig. 9. (a) Bi-beam generated by our two-stage algorithm in 19-element CBC system. (b) Emission plane in 19-element CBC system. (c) Bi-beam generated by our two-stage algorithm in 127-element CBC system. (d) Emission plane in 127-element CBC system.

Another interesting application is that we can control the power of these two beams by adjusting $k_0$ and $r_0$ in (31), so as to generate unbalanced Bi-beam. The results are shown in the Fig. 10. However, when unbalanced Bi-beam is generated, the positions of the beams in the far field are different from the ones of balanced Bi-beam. We will also further study the relationship between the power ratio and the relative position of multiple beams in the future work.

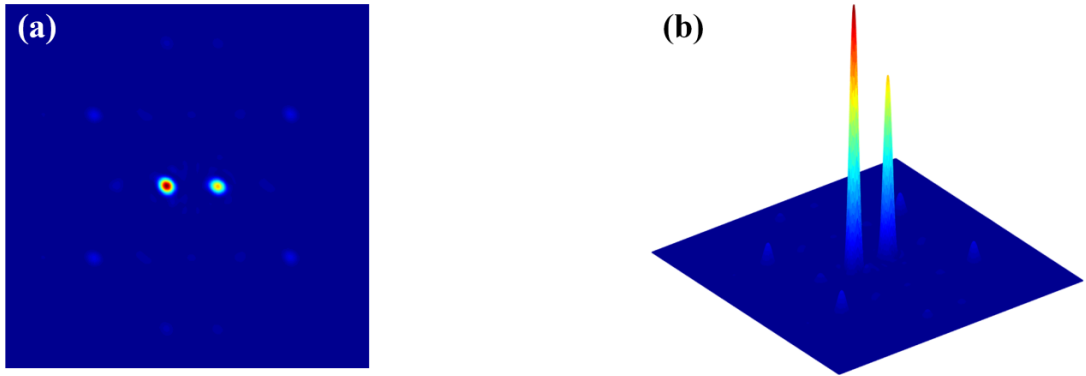

Fig. 10. Unbalanced Bi-beam could be generated by adjusting parameters in (26).

## 5. Summary

In this paper, we prove that phase ambiguity will occur in coherent beam combining (CBC) system with centrosymmetric distribution of subapertures. As far as we know, we are the first to obtain all solutions of phase ambiguity in multi-aperture CBC system through theoretical derivation: if two groups of piston aberrations generate the same far-field image, they are equal or rotationally conjugate to each other. To solve this problem, we propose a method by applying asymmetric phase modulation without adding additional optical devices. Simulation results have proved our conclusions. In addition, we have designed a novel two-stage algorithm of Bi-beam generation. We believe that our work can not only help to theoretically analyze the corresponding relationship between piston aberration and far-field image, but also extended applications of coherent beam combining.

## Appendix

*1. Phase redundancy*

We find that when adding a common piston aberration $\rho$ to all subapertures in $\Psi = \{\psi_n\}$, the value of equation (9) remains unchanged. We call the multi solution problem caused by this *phase redundancy*. As shown in Fig. 11, $\Psi' = \{\psi_n + \rho\}$ generate the same far-field image as $\Psi = \{\psi_n\}$. When the value of $\rho$ changes, there exists an infinite number of groups of piston aberrations that may generate the same far field. After selecting a subaperture's piston aberration as the reference (The selection of subaperture does not affect the conclusion. In this paper, we select subaperture0 as the reference.), we subtract the reference piston aberration from the initial piston aberration and obtain the relative phase. Thus, the above infinite cases will be transformed into a unique expression $\Phi = \{\varphi_n\}$, where $\varphi_n = \psi_n - \psi_0$. Because of

$\varphi_0 = \psi_0 - \psi_0 \equiv 0$, $\varphi_0$ can usually be omitted. That is, $\Phi = \{\varphi_n\}$ contains only the relative piston aberrations of N-1 subapertures other than the reference one (subaperture0). Unless otherwise specified, "piston aberration" in the following text refers to the relative piston aberration.

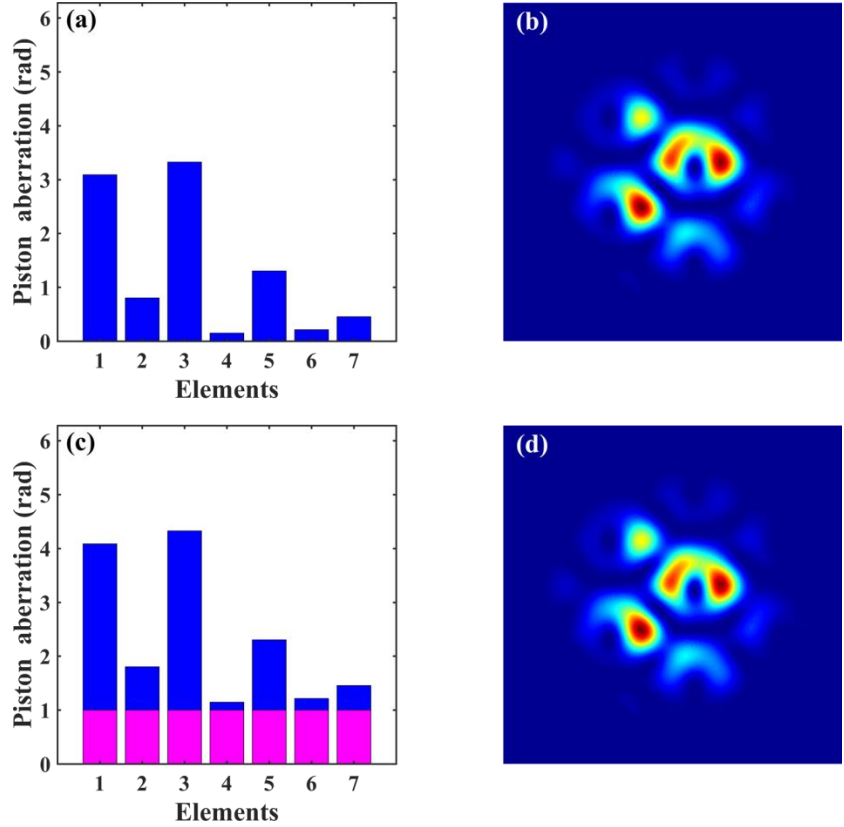

Fig. 11. After adding a common piston aberration, the far-field remains unchanged.

*2. Two-stage algorithm*

Our Bi-beam generation algorithm consists of two stages. In the first stage, we obtain the coarse position of Bi-beam in the far field. We may ask whether this step is necessary. The question is whether we can choose the positions of Bi-beam arbitrarily, and then generate balanced Bi-beam by maximizing the energy in these areas (which means skipping the stage 1). Our answer is No. We will show three negative consequences resulted by this approach.

(1) Strip spot

If we select the adjacent areas as the position where the Bi-beam appears, a strip spot may be generated after optimization (which is shown in Fig. 12). At this time, the primary and secondary beams are not separated. So it is failed to obtain multiple beams.

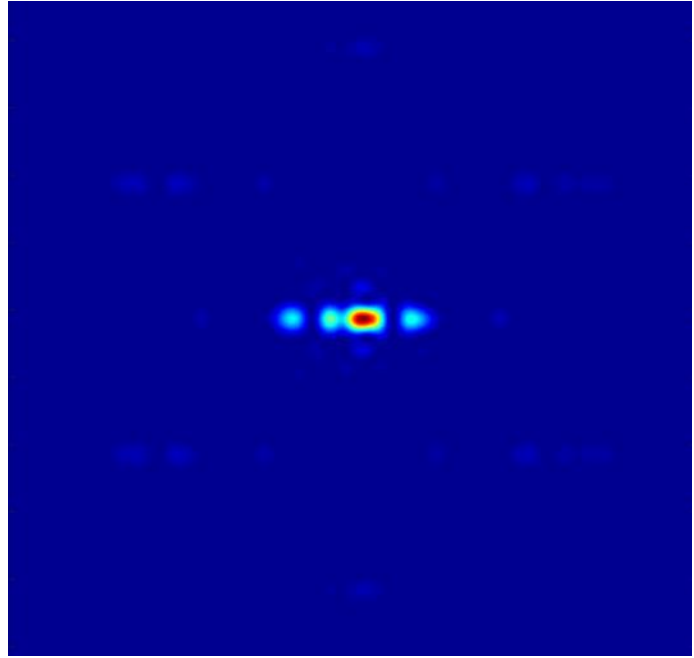

Fig. 12. A strip spot may be generated in adjacent grids

(2) Unbalanced Bi-Beam

If we select the asymmetrically arranged areas, we will not be able to generate balanced Bi-beam. The result is shown in Fig. 13, which will limit its application.

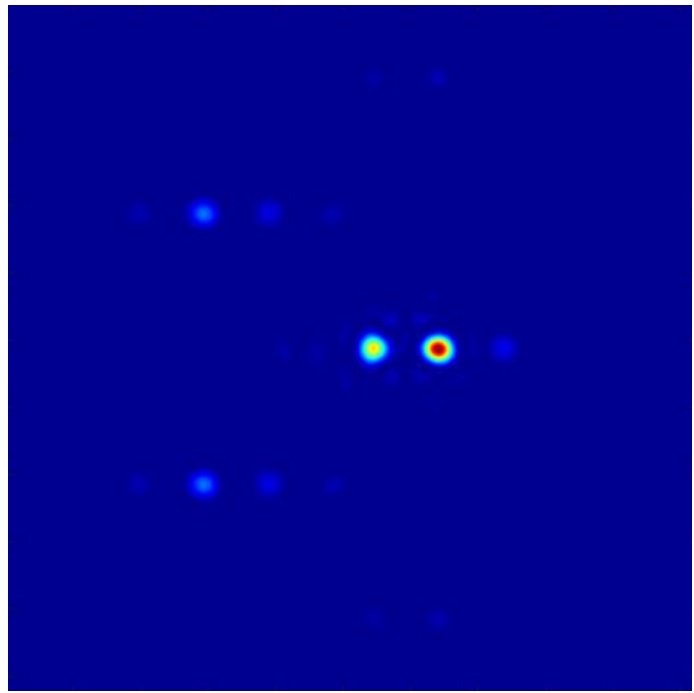

Fig. 13. Unbalanced Bi-beam may be generated in asymmetrically arranged areas.

(3) Higher sidelobe energy

If we randomly select the symmetrically arranged areas which are not adjacent to each other, we may generate separated and balanced Bi-Beam. However, compared with the Bi-Beam obtained by our two-stage algorithm, it usually has higher sidelobe energy (which is shown in Fig. 14).

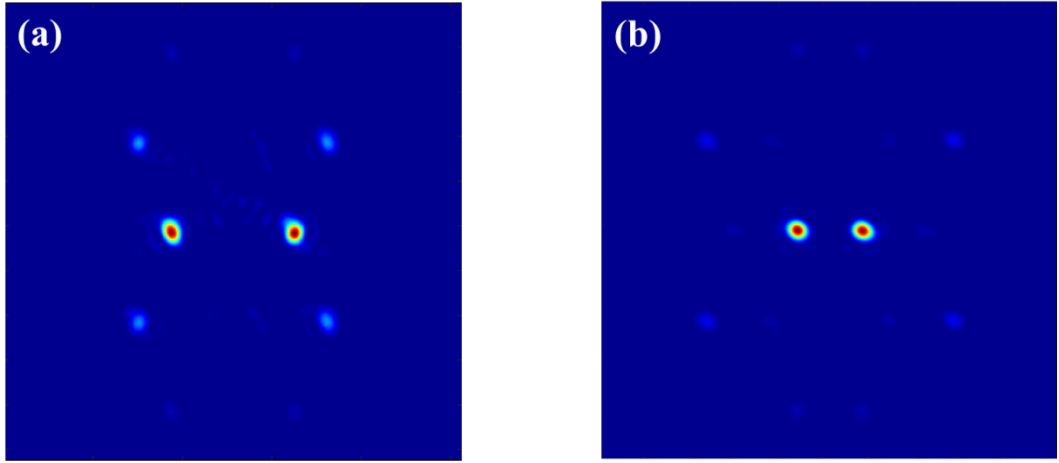

Fig. 14. (a) Bi-beam generated by randomly selecting the symmetrically arranged areas. (b) Bi-beam generated by our two-stage algorithm.

Thus, randomly selecting the target areas and optimizing the beams' energy in these areas is not conducive to obtaining separate, balanced and energy-concentrated Bi-beam, which indicates that our two-stage algorithm is beneficial for improving the quality of Bi-beam.

**Funding**

National Natural Science Foundation of China (62175241, 62005286).

**Disclosures**

†Haolong Jia and Jing Zuo are co-first authors. The authors declare no conflicts of interest.